\documentclass[pdflatex,sn-mathphys-num]{sn-jnl}% Math and Physical Sciences Numbered Reference Style 
\usepackage{graphicx}%
\usepackage{multirow}%
\usepackage{amsmath,amssymb,amsfonts}%
\usepackage{amsthm}%
\usepackage{mathrsfs}%
\usepackage{adjustbox}%%
\usepackage{array}%
\usepackage[title]{appendix}%
\usepackage{xcolor}%
\usepackage{textcomp}%
\usepackage{manyfoot}%
\usepackage{booktabs}%
\usepackage{algorithm}%
\usepackage{algorithmicx}%
\usepackage{algpseudocode}%
\usepackage{listings}
\usepackage{hyperref}

\theoremstyle{thmstyleone}%
\theoremstyle{thmstyletwo}%

\theoremstyle{thmstylethree}%

\begin{document}

\title[Using virtual reality to enhance mobility, safety, and equity for persons with vision loss in urban environments]{Using virtual reality to enhance mobility, safety, and equity for persons with vision loss in urban environments}


\author[1,2]{\fnm{Fabiana Sofia} \sur{Ricci}}\email{fsr8343@nyu.edu}

\author[3]{\fnm{Charles K.} \sur{Ukegbu}}\email{cukegbu@dot.nyc.gov}

\author[3]{\fnm{Anne} \sur{Krassner}}\email{akrassner@dot.nyc.gov}

\author[3]{\fnm{Sanjukta} \sur{Hazarika}}\email{shazarika@dot.nyc.gov}

\author[3]{\fnm{Jade} \sur{White}}\email{jade32495@gmail.com}

\author[1,2,4]{\fnm{Maurizio} \sur{Porfiri}}\email{mporfiri@nyu.edu}

\author*[1,4,5,6]{\fnm{John-Ross} \sur{Rizzo}}\email{JohnRoss.Rizzo@nyulangone.org}

\affil[1]{\orgdiv{Department of Biomedical Engineering}, \orgname{New York University Tandon School of Engineering}, \orgaddress{\street{Six MetroTech Center}, \city{Brooklyn, NY}, \postcode{11201}, \state{New York}, \country{USA}}}

\affil[2]{\orgdiv{Center for Urban Science and Progress}, \orgname{New York University Tandon School of Engineering}, \orgaddress{\street{370 Jay Street}, \city{Brooklyn, NY}, \postcode{11201}, \state{New York}, \country{USA}}}

\affil[3]{\orgdiv{Transportation Planning \& Management Division}, \orgname{New York City Department of Transportation}, \orgaddress{\street{55 Water Street}, \city{New York}, \postcode{10041}, \state{NY}, \country{USA}}}

\affil[4]{\orgdiv{Department of Mechanical and Aerospace Engineering}, \orgname{New York University Tandon School of Engineering}, \orgaddress{\street{Six MetroTech Center}, \city{Brooklyn, NY}, \postcode{11201}, \state{New York}, \country{USA}}}

\affil[5]{\orgdiv{Department of Rehabilitation Medicine}, \orgname{New York University Langone Health}, \orgaddress{\street{ 240 East 38th
Street}, \city{New York}, \postcode{10016}, \state{NY}, \country{USA}}}

\affil[6]{\orgdiv{Department of Neurology}, \orgname{New York University Langone Health}, \orgaddress{\street{240 East 38th Street}, \city{New York}, \postcode{10016}, \state{NY}, \country{USA}}}


\abstract{This study explores the use of virtual reality (VR) as an innovative tool to enhance awareness, acceptance, and understanding of accessibility for persons with vision loss (VL). Through a VR-based workshop developed in collaboration with New York City's Department Of Transportation, participants experienced immersive simulations of VL and and related immersive mobility challenges. The methodology included pre- and post-intervention questionnaires, assessing changes in participants’ knowledge, confidence, and perception. Participants included urban planners, designers, and architects. Results showed a significant increase in awareness of VL-related challenges that affect design guidelines, as well as improved confidence in addressing such challenges. Participants also expressed strong support for VR as a pedagogical tool, noting its potential for reshaping professional practices, improving capacity building, and enhancing inclusive design. The study demonstrates the effectiveness of VR as an experiential learning platform, fostering empathy and a long-term commitment to integrating VL considerations into urban design. These findings highlight the transformative potential of VR in advancing equity and accessibility in urban environments.}




\maketitle

\section{Introduction}\label{Introduction}
Urban environments are bustling hubs of activity, filled with diverse spaces and pathways. For persons with vision loss (VL), navigating through these spaces presents a multitude of challenges that often go unnoticed by the sighted population \cite{campisi2021evaluation,el2021systematic}. In fact, for someone with a VL, moving through city streets, squares, and sidewalks can be fraught with obstacles and difficulties that can affect not only physical movement but also emotional well-being \cite{campisi2021evaluation,el2021systematic}.

One of the most significant hurdles faced by persons with VL is the presence of physical barriers and infrastructural challenges \cite{mountapmbeme2022addressing,park2022towards,muller2022traveling}. Uneven pavement, poorly maintained sidewalks, and obstacles such as parked cars or street furniture can disrupt the flow of movement and pose safety risks. Stairs and curbs, in particular, represent significant barriers, requiring alternative routes or assistance to navigate \cite{mountapmbeme2022addressing,park2022towards,muller2022traveling}. Moreover, the geometric and material conditions of public spaces play a crucial role in usability \cite{upadhyay2022inclusive}. Factors such as the width and slope of pedestrian pathways, the presence and arrangement of seating, and the stability and roughness of pavement surface can either facilitate or hinder safe navigation \cite{kapsalis2024disabled}. Ensuring that these elements are designed with consideration for all users is vital for creating inclusive urban spaces.

Urban environments are rich in sensory stimuli, which can be both an asset and a challenge for persons with VL \cite{little2020experiential}. Auditory cues such as the sound of traffic or pedestrian signals are essential for navigation. However, high levels of ambient noise can interfere with the ability to detect these important sounds, leading to disorientation and confusion \cite{kayukawa2020guiding,may2020spotlights,gupta2020towards}. Visual cues, though less directly useful for those with severe VL, still play a role for persons with low vision. Factors like color contrast, lighting, and the placement of tactile indicators are critical for enhancing the legibility of spaces \cite{wabinski2024evaluation,mattsson2020improved,hauck2022investigation}. Reflectance levels, positioning of lights, and clear, consistent tactile markings can significantly improve the navigability of urban areas for persons with VL \cite{wabinski2024evaluation,mattsson2020improved,hauck2022investigation}.

Beyond physical obstacles, the emotional impact of navigating urban spaces on persons with VL cannot be overstated. Feelings of anxiety, frustration, and insecurity often accompany the experience of encountering barriers and uncertainties along travel routes \cite{ibe2023take,cushley2023unseen}. The fear of getting lost, tripping, or facing unexpected hazards adds layers of stress to an already challenging journey \cite{ibe2023take,cushley2023unseen}. The stress and anxiety caused by navigating unfamiliar or poorly designed spaces can then deter persons with VL from fully participating in urban life \cite{rizzo2023global,conboy2022lived,moles2022travel}. This exclusion not only affects their emotional well-being but also limits their opportunities for social interaction, community engagement and professional development \cite{rizzo2023global,conboy2022lived,moles2022travel}.

While accessibility guidelines often exist, their implementation varies widely, leading to inconsistencies in accessibility across urban areas \cite{nicoletti2023disadvantaged,levine2023approaching,handy2020accessibility}. What might seem like a simple walkable path to a sighted person can become a maze of challenges to persons with VL. Spaces that lack tactile paving, auditory signals at crossings, or clear signage can be especially disorienting \cite{annakin2023can,scalvini2023outdoor}. This underscores the necessity of inclusive design principles, such as Universal Design, which advocates for creating environments that are usable and meaningful for all \cite{pinna2021literature,pinna2020beyond,steinfeld2023space}. By designing public spaces that are accessible and accommodating to persons with VL, cities can promote social inclusion and enhance quality of life for all residents.

\section{Methodology}
To promote accessibility and the redesign of urban spaces based on Universal Design principles, we organized the InSight workshop in collaboration with the New York University Langone Health (NYULH) and the New York City (NYC) Department of Transportation (DOT) aimed at highlighting perceptual difficulties and related mobility challenges faced by persons with VL while navigating complex urban environments. The concept was that deeper understanding should encourage solutions that make cities more accessible. The workshop involved members of the NYC DOT who experienced simulations of different VLs and traveled through virtual urban environments resembling NYC. Participants were also asked to fill out a short questionnaire before and after the intervention to evaluate the effectiveness of the workshop. Finally, a focus group and follow-up interviews with members from various departments of the DOT were conducted to foster open conversations about accessibility issues in NYC for people with disabilities. These discussions also covered current projects aimed at improving inclusivity and safety, and potential interventions for further enhancements. The following subsections describe the workshop setup, procedures, and evaluation of its impact through participant questionnaires and post-workshop discussions.

\subsection{Virtual reality design}
The immersive virtual urban environment was designed using Unity version 2019.4.28f1 and the virtual reality (VR) headset was the Oculus Quest with handheld controllers. Our VR platform aims at providing the DOT with a tool that accurately reproduces critical areas of NYC. The goal is to raise awareness among personnel about the navigation challenges and risks experienced by persons with VL in urban environments. By engaging with this immersive simulation, DOT personnel can gain a deeper understanding of and sensitivity towards the needs of persons with VL and foster efforts to promote the implementation of design interventions that improve accessibility and safety of urban environments.
\begin{figure}[htb]
\centering
\begin{tabular}{cc}
    (a) & (b) \\
   \includegraphics[width=0.45\textwidth]{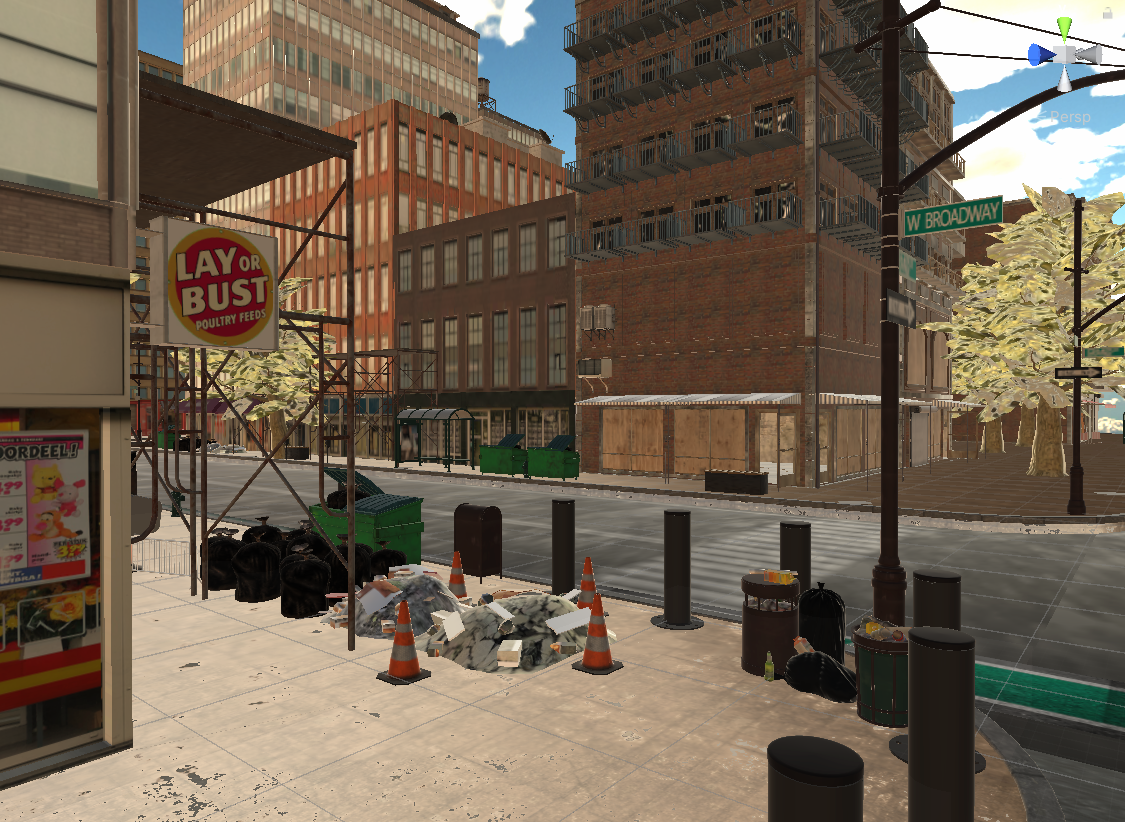}  & \includegraphics[width=0.50\textwidth]{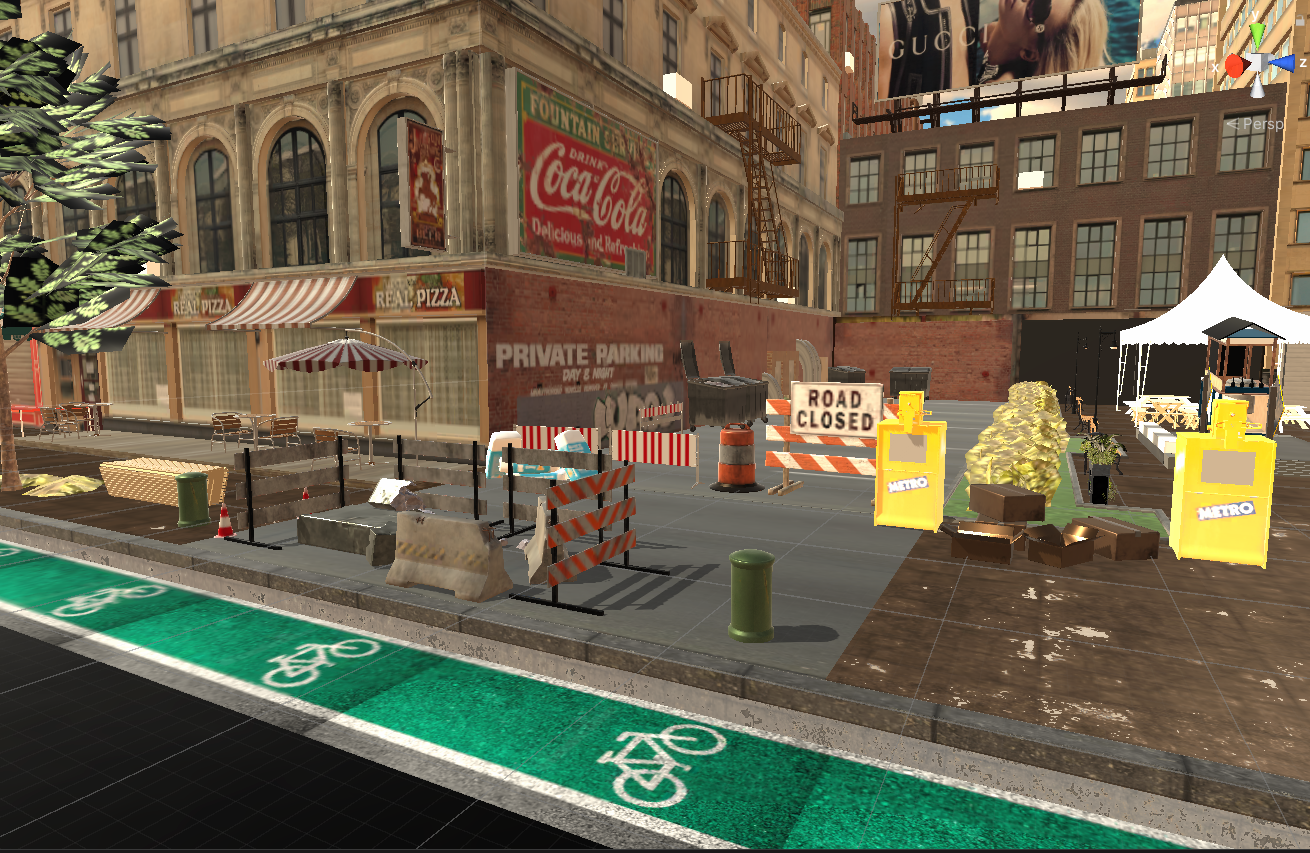} \\
\end{tabular}
\begin{tabular}{c}
    (c) \\  \includegraphics[width=0.49\textwidth]{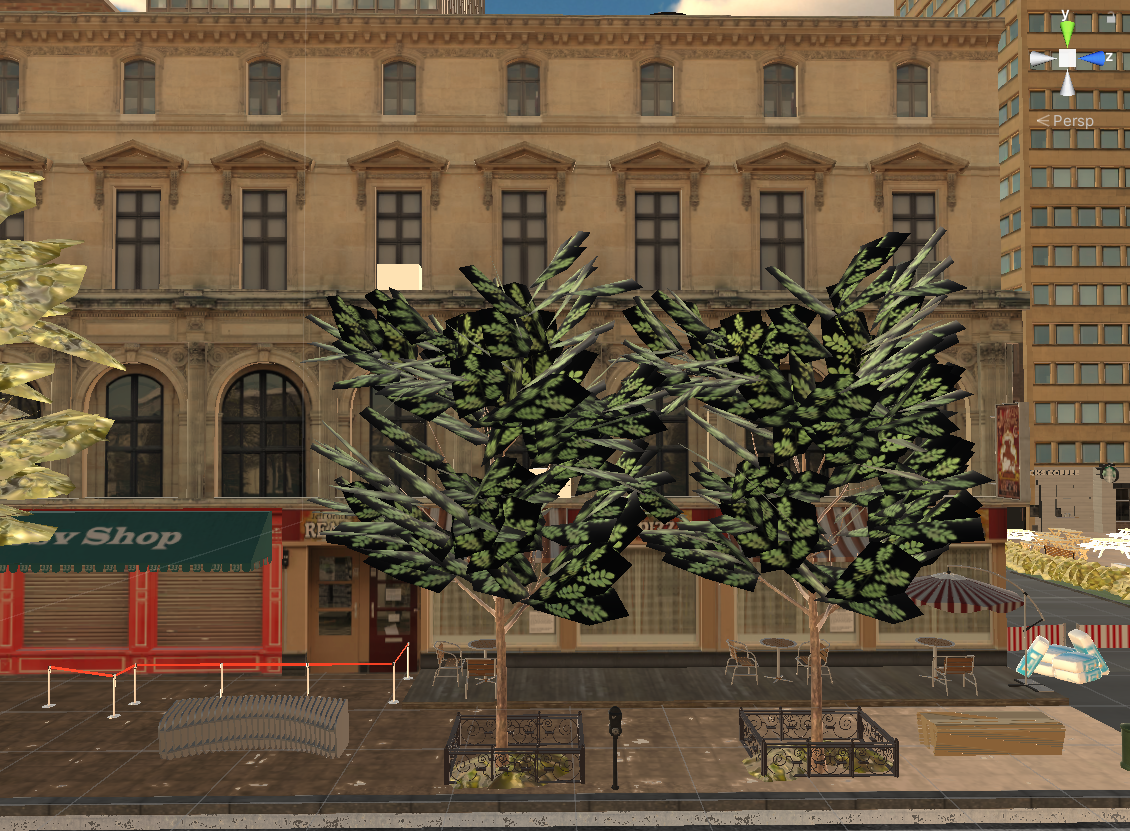} \\
\end{tabular}
\caption{Examples of obstacles included in the virtual urban scenario: (a) traffic cones, trash, and bollards, (b) construction site, and (c) crowd control barriers and benches .}   
\label{fig:fig1}
\end{figure}

For the design of the virtual environments, we drew inspiration from real-life urban areas in NYC that are particularly challenging for persons with VL. The environments were created to realistically replicate the complex and often hazardous navigation challenges faced by persons with VL in an urban setting. The setup included a variety of obstacles and hazards that persons with VL may encounter such as bollards, trash, construction sites, scaffolding, barricades, crowd control barriers, and benches, all strategically placed to create misalignments and unexpected threats (see Fig. \ref{fig:fig1}). 
\begin{figure}[!h]
\centering
\begin{tabular}{cc}
    (a) & (b) \\
\includegraphics[width=0.52\textwidth]{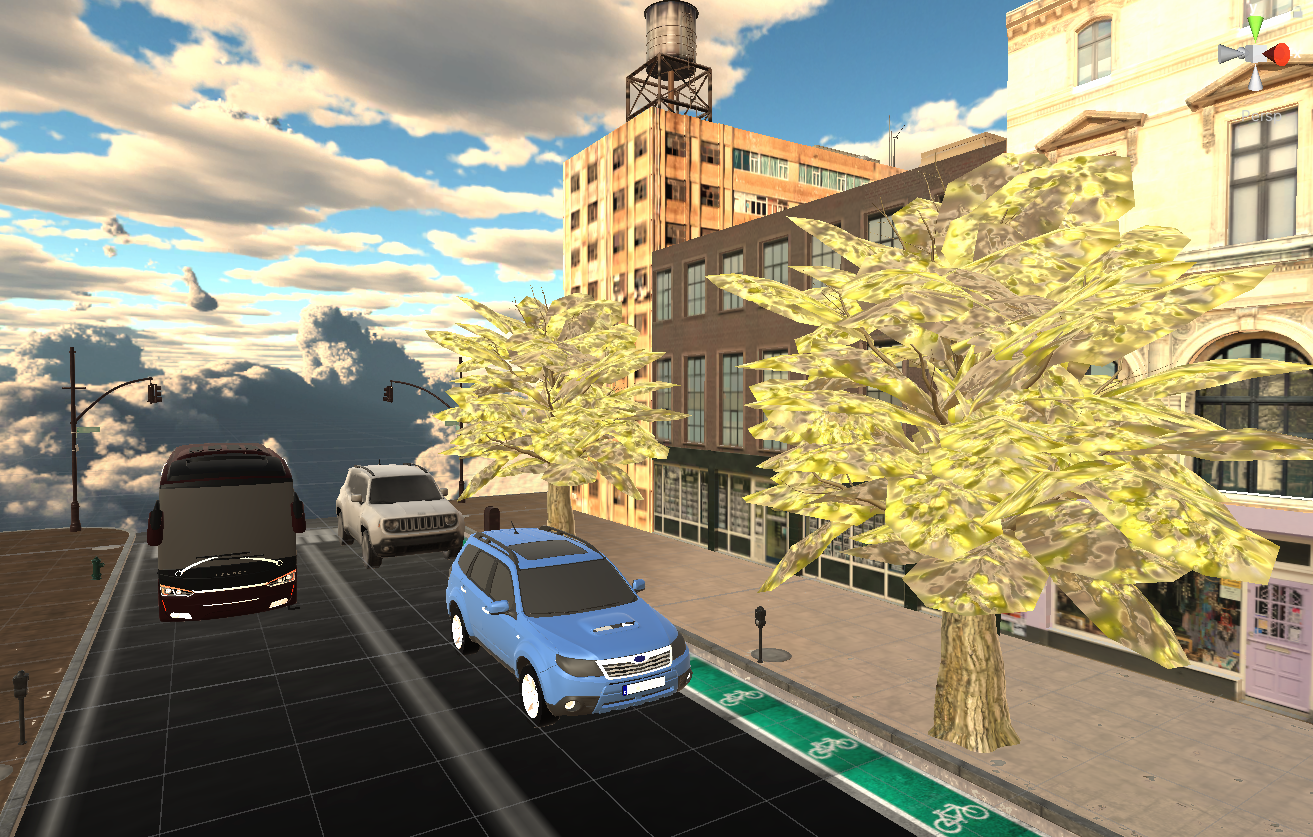}  & \includegraphics[width=0.45\textwidth]{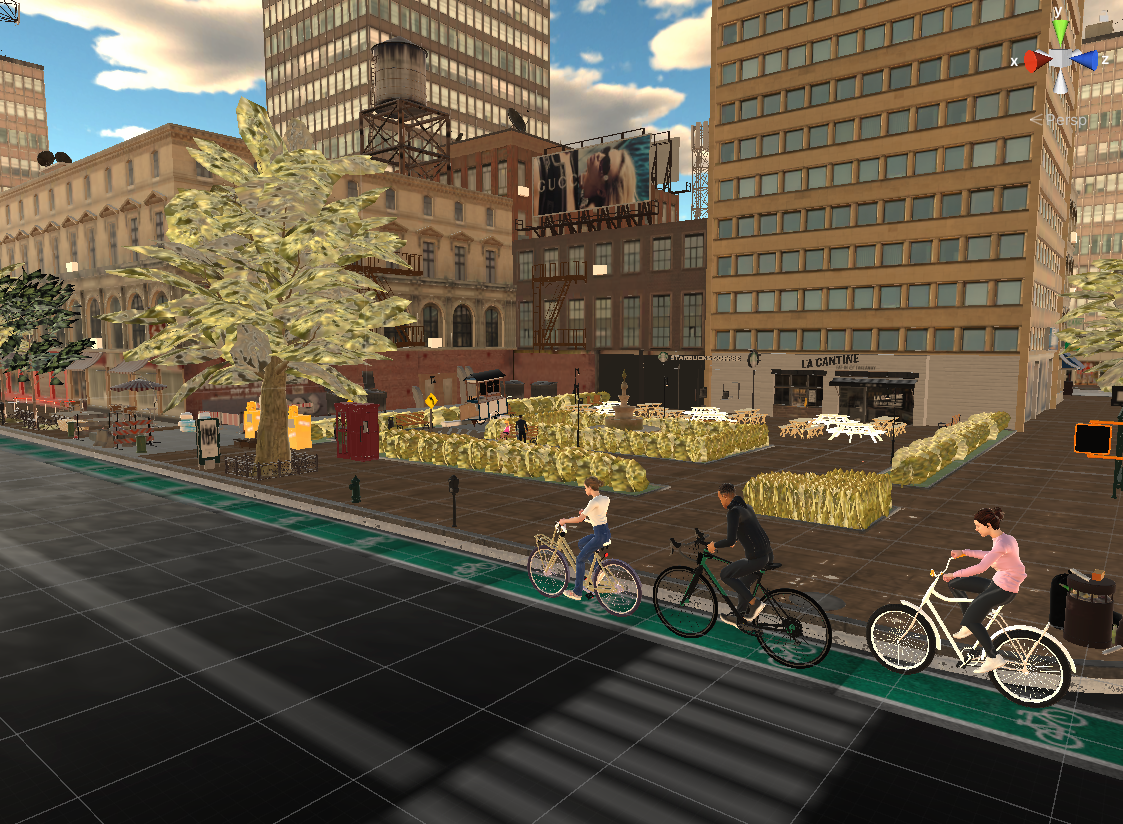} \\
\end{tabular}
\begin{tabular}{c}
    (c) \\  \includegraphics[width=0.45\textwidth]{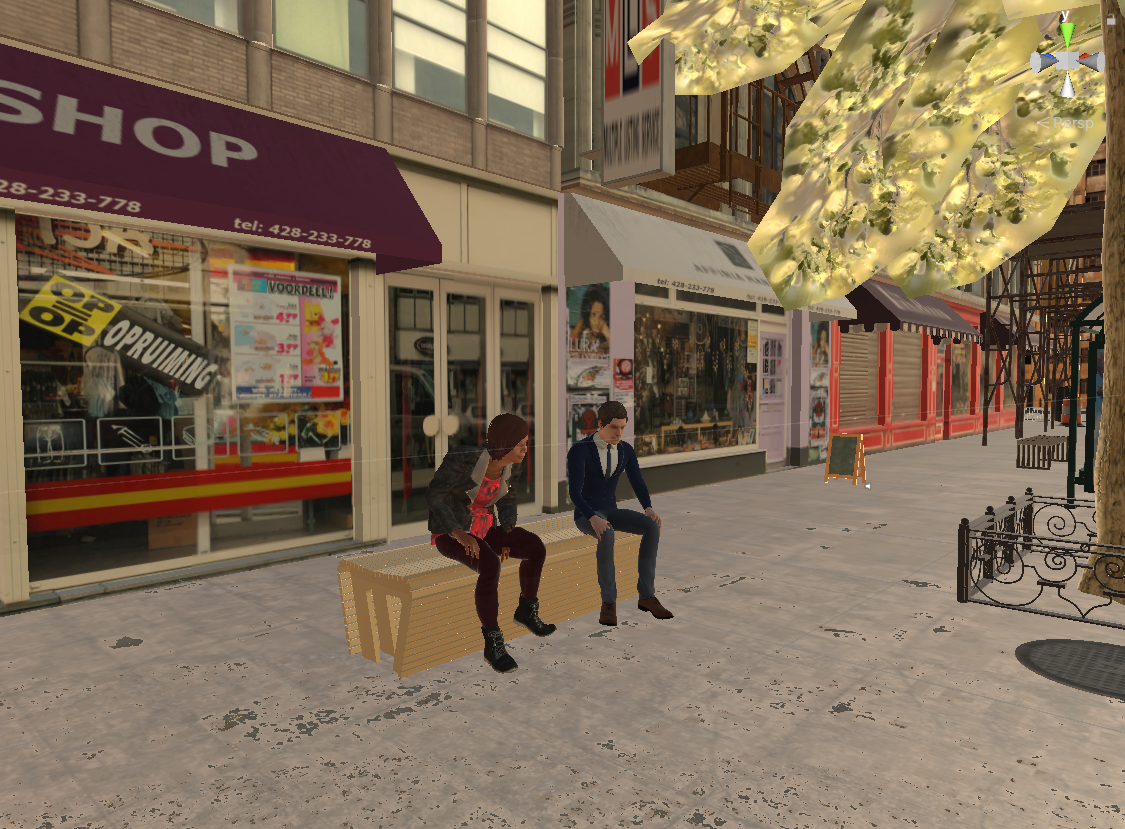} \\
\end{tabular}
\caption{Examples of animations included in the virtual urban scenario: (a) car traffic, (b) cyclists, and (c) characters.}   
\label{fig:fig2}
\end{figure}
To enhance the immersiveness and realism of the experience, we incorporated animated characters, regulated car traffic by traffic lights, and a bicycle lane with animated cyclists (see Fig. \ref{fig:fig2}). 
\begin{figure}[htb]
\centering
\begin{tabular}{c} \\
\includegraphics[width=0.49\textwidth]{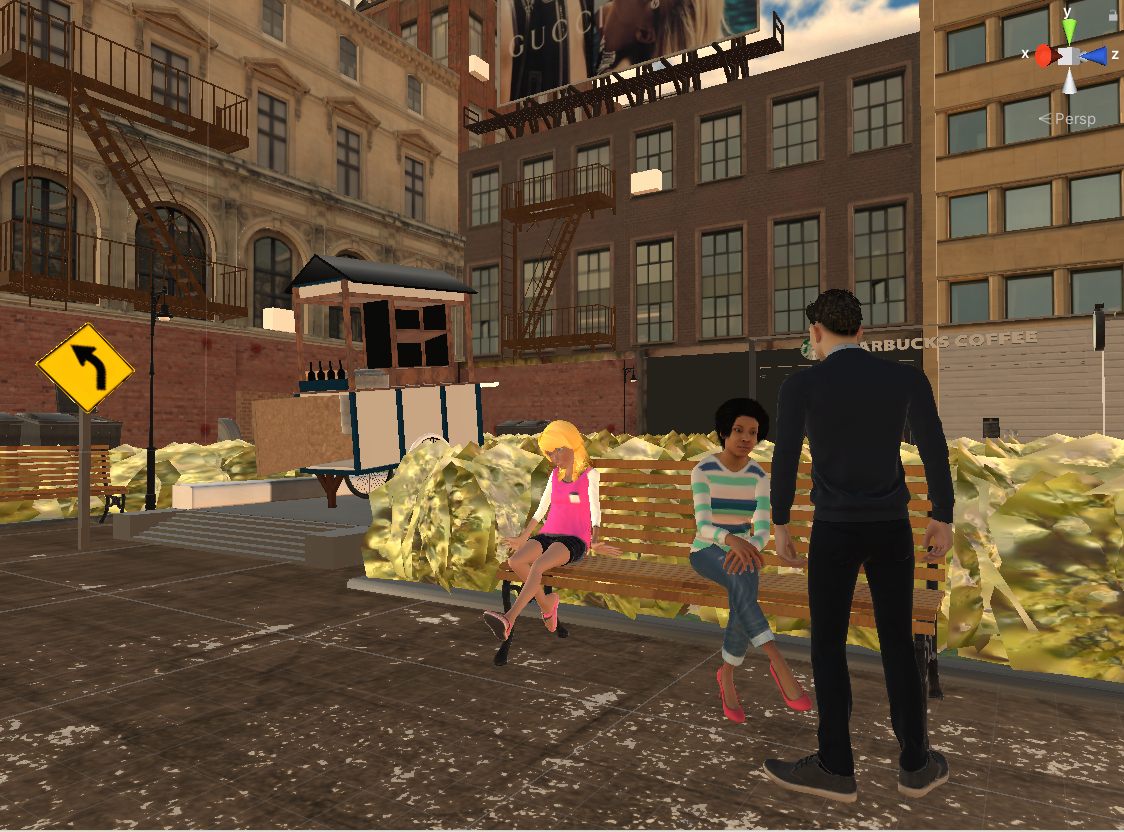}  \\
\end{tabular}
\caption{Example of audio cue included in the virtual urban scenario: animated characters sitting on a bench.}   
\label{fig:fig3}
\end{figure}
Specialized audio cues were matched to the user's egocentric heading. For instance, if a user walked close to a bench on their right side where animated characters were sitting, the sound of their voices would be heard in the user's right ear (Fig. \ref{fig:fig3}). When facing the sidewalk, the street sounds were heard in both ears. Additionally, global environmental sounds were implemented to create a more engaging and interactive experience, offering crucial auditory cues about the surroundings, including those related to potential distractions. Finally, audio guides were provided throughout the experience to assist users in avoiding obstacles and to offer directional instructions.

\subsection{vision loss simulation}
In our platform, we simulated various types of VL to home a comprehensive understanding of the challenges faced by persons with such conditions. Simulated VLs included the three most widespread eye pathologies in the United States such as age-related macular degeneration (AMD), diabetic retinopathy (DR), and glaucoma, each at three stages: mild, moderate, and severe \cite{wittenborn2023prevalence}.

\begin{figure}[!h]
\centering
\begin{tabular}{cc}
    (a) & (b) \\
\includegraphics[width=0.478\textwidth]{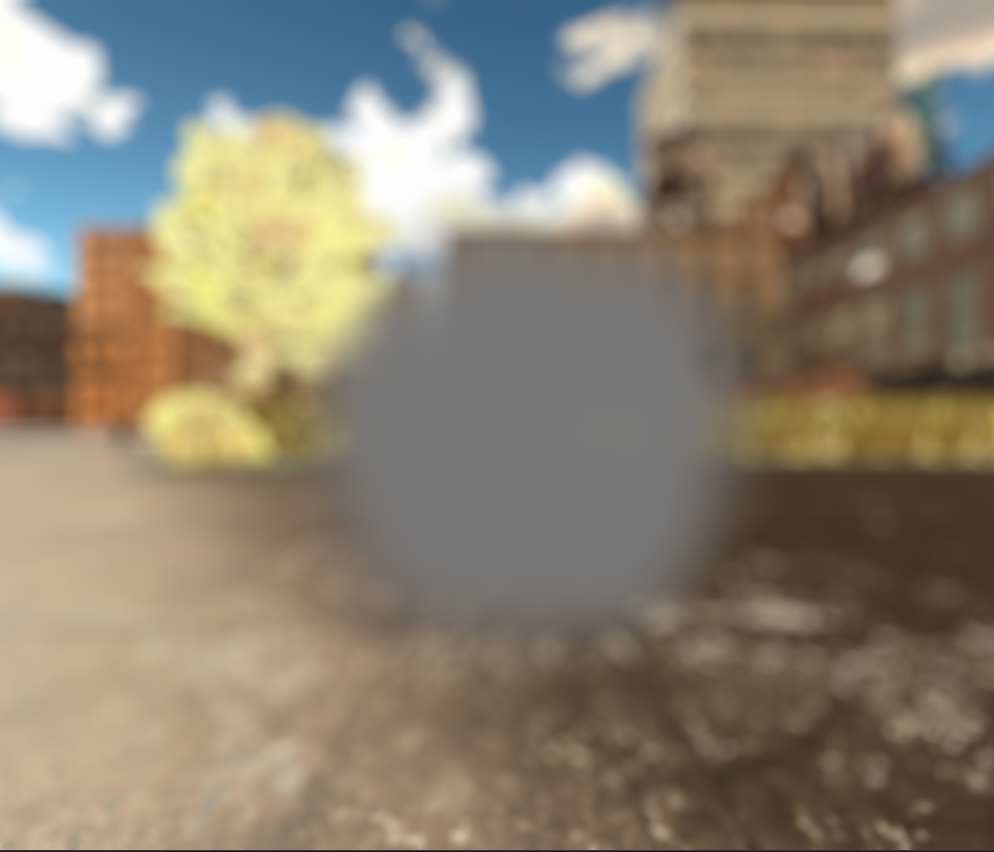}  & \includegraphics[width=0.49\textwidth]{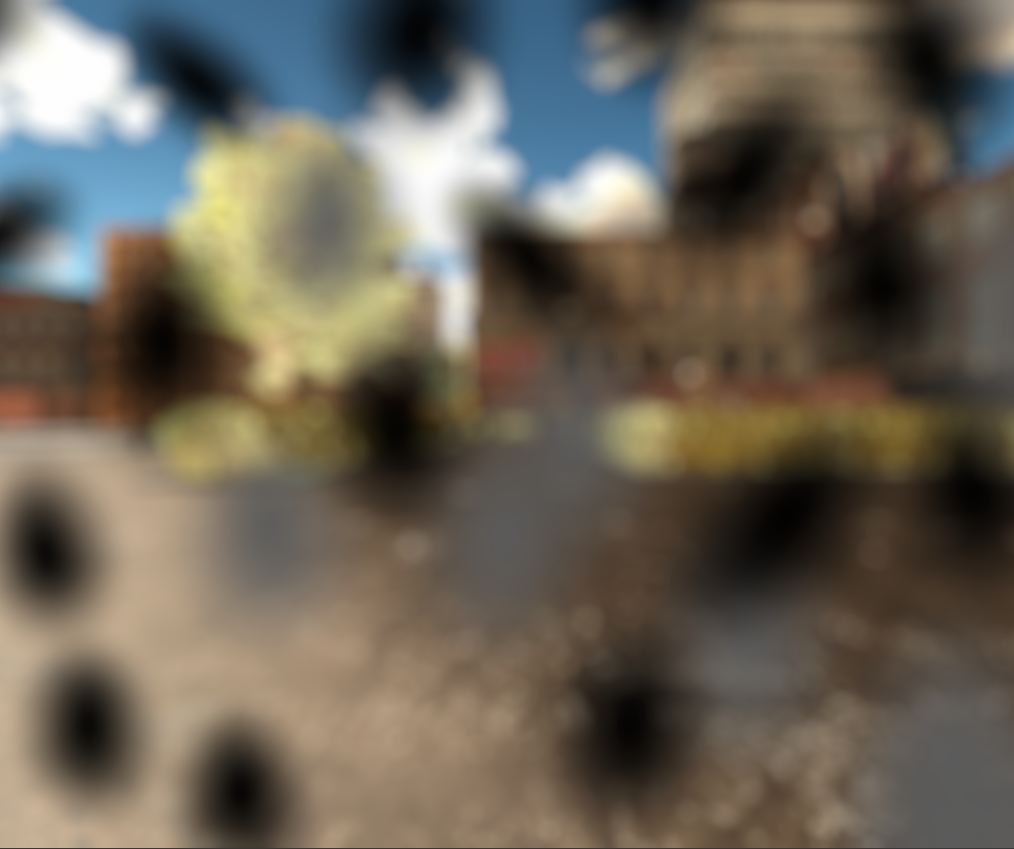} \\
\end{tabular}
\begin{tabular}{c}
    (c) \\  \includegraphics[width=0.55\textwidth]{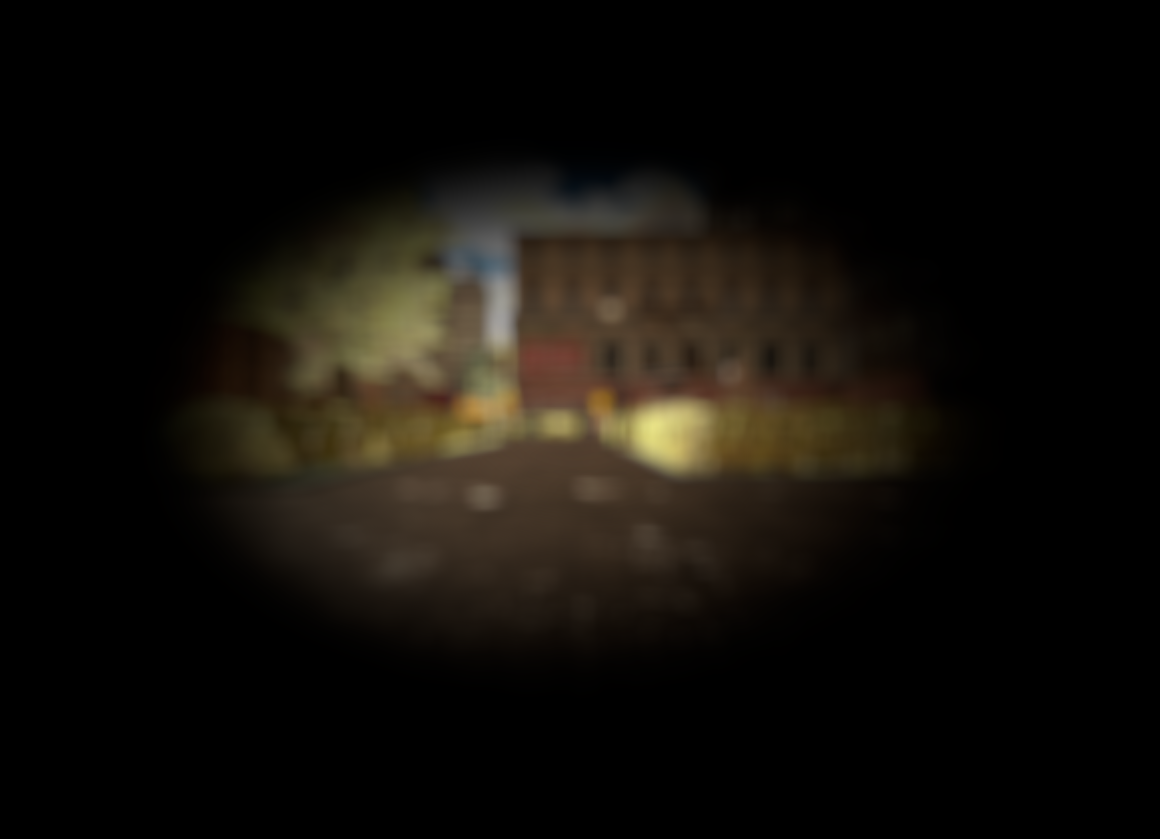} \\
\end{tabular}
\caption{Examples of simulated visual disabilities at severe stage: (a) age-related macular degeneration, (b) diabetic retinopathy, and (c) glaucoma.}   
\label{fig:fig4}
\end{figure}

AMD is a common eye condition that primarily affects individuals over the age of 50 \cite{fleckenstein2024age}. It is caused by the deterioration of the central portion of the retina, known as the macula, which is responsible for sharp, central vision \cite{amini2023mechanistic}. Characteristic symptoms of AMD include blurriness, dark areas, or distortion in the center of the visual field (see Fig. \ref{fig:fig4}(a)) \cite{amini2023mechanistic}. AMD particularly affects the ability to discern fine details, making navigation in unfamiliar environments demanding \cite{kallstrand2024perpetuating}.

DR is a complication of diabetes that affects the blood vessels of the retina \cite{srivastava2024diabetes}. It is caused by high blood sugar levels damaging the retinal blood vessels, leading to swelling, leakage, and eventually the closure of these vessels \cite{srivastava2024diabetes}. Symptoms of DR include floaters, blurriness, and dark areas of vision (see Fig. \ref{fig:fig4}(b)). In advanced stages, it can lead to a significant reduction in visual acuity, depth perception, and color perception, impacting the ability to detect obstacles and hazards while navigating \cite{srivastava2024diabetes,cushley2023role}. 

Glaucoma is a group of eye conditions that damage the optic nerve, often due to abnormally high pressure in the eye \cite{srivastav2024glaucoma}. It is a leading cause of blindness for people over the age of 60 \cite{srivastav2024glaucoma}. Glaucoma progresses slowly and in its mild stage, symptoms are often not noticeable \cite{sahu2024review}. However, as it progresses to moderate and severe stages, it can cause loss of peripheral vision, tunnel vision, and eventually total blindness if left untreated (see Fig. \ref{fig:fig4}(c)) \cite{sahu2024review}. Glaucoma symptoms may impact the ability of detecting obstacles from the sides \cite{deemer2023approaching}.

The simulation of the three VLs was achieved through a combination of shader effects and culling masks within the Unity environment. Shaders, which are scripts capable of altering light behavior on object surfaces, were utilized to manipulate light, darkness, and color to replicate VLs realistically. Culling masks, on the other hand, allowed for the selective rendering of specific parts of the scene to generate targeted visual effects. To simulate AMD, we employed Gaussian blur and distortion shaders to mimic the blurriness and distortion experienced by persons with this condition. Gray spots that obscure the central visual field were integrated using a culling mask, replicating the hallmark symptoms of AMD. For DR, we utilized Gaussian blur shaders to create a blurred effect across the visual field, while dark spots characteristic of DR were scattered throughout using a culling mask. This combination accurately reproduced the visual disturbances associated with DR. Similarly, symptoms of glaucoma were simulated by applying Gaussian blur shaders to induce blurriness, particularly in the peripheral vision. The culling mask was used to compromise the peripheral visual field, replicating the tunnel vision often experienced by persons with glaucoma.

\subsection{InSight workshop}
The NYU-DOT InSight workshop was conducted at the NYU Tandon School of Engineering, located at 370 Jay Street, Brooklyn, NY 11201. The event took place on the 12th floor, in seminar room \#1201, with 11 members of the NYC DOT participating spanning leadership teams from the Transportation Planning and Management Division, including Geometric Design, Traffic Engineering and Planning, Traffic Control and Engineering, Regional and Strategic Planning, Transit Development, the Office of Livable Spaces (which focuses on bicycles, pedestrians, and the public realm), and the Research Implementation and Safety unit. 

\begin{figure}[!h]
\centering
\begin{tabular}{cc}
    (a) & (b) \\
\includegraphics[width=0.49\textwidth]{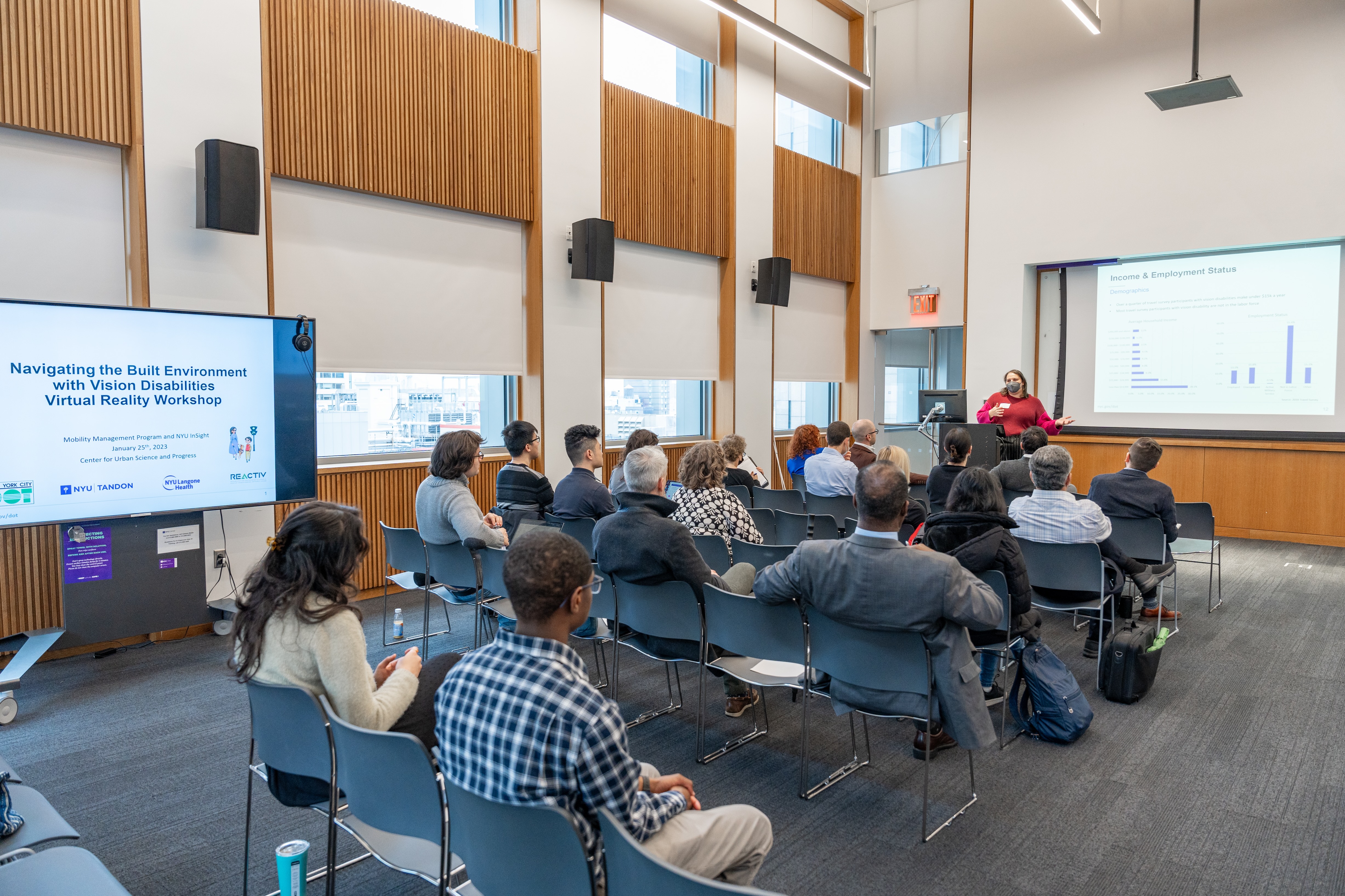}  & \includegraphics[width=0.493\textwidth]{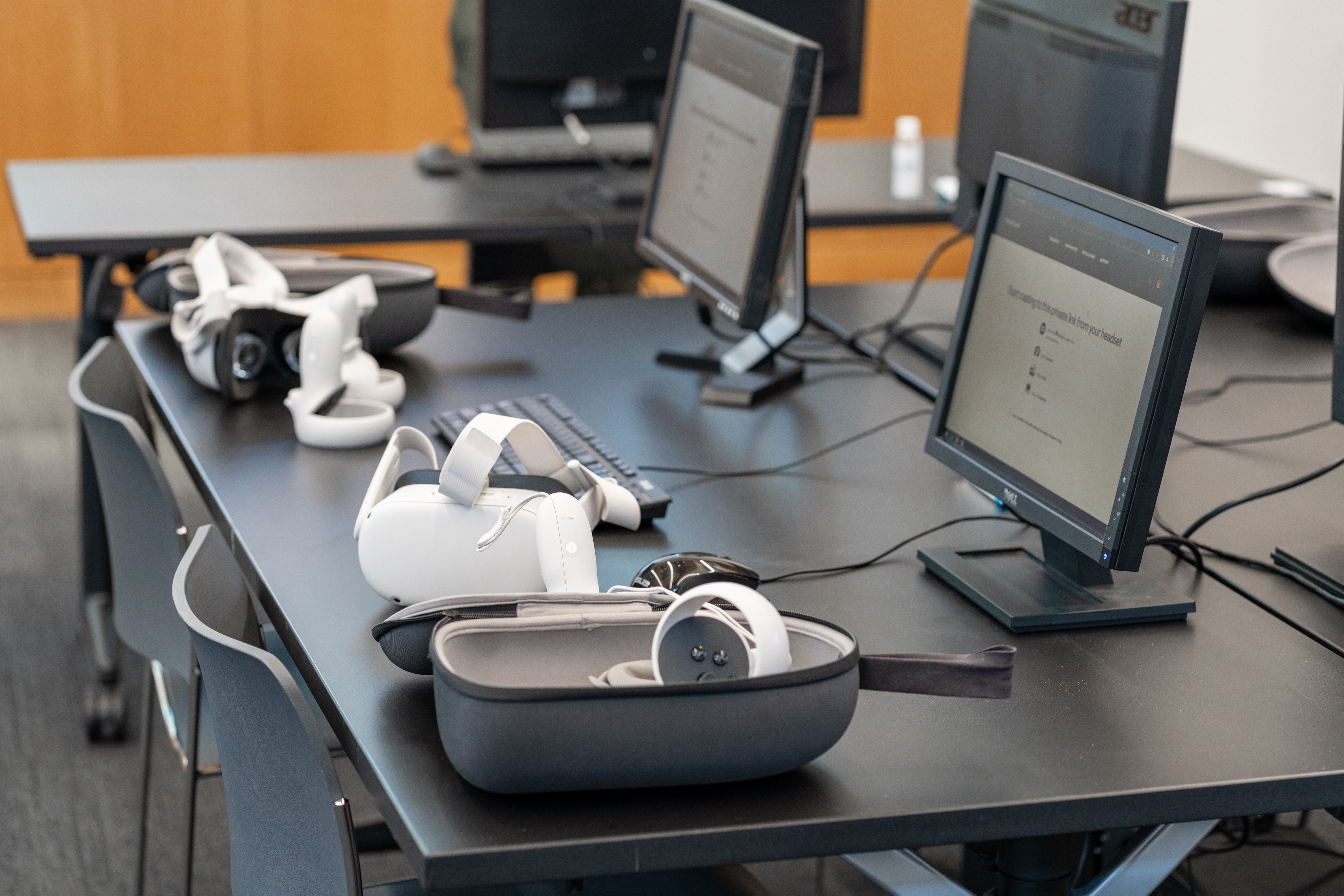} \\
\end{tabular}
\caption{Set-up of the workshop: (a) project introduction and (b) VR headsets and PCs prepared for participant use}   
\label{fig:fig5}
\end{figure}

The workshop commenced with an overview of the overall project aimed at improving the safe and independent mobility of persons with VL (see Fig. \ref{fig:fig5}a). Following this introduction, detailed information was provided about the upcoming VR experience, including the environmental design, the simulated VLs, and instructions on how to use the VR headset and controllers. Before and after the VR experience, participants were asked to fill out a pre- and post-intervention questionnaire.

\begin{figure}[!h]
\centering
\begin{tabular}{cc}
    (a) & (b) \\
\includegraphics[width=0.49\textwidth]{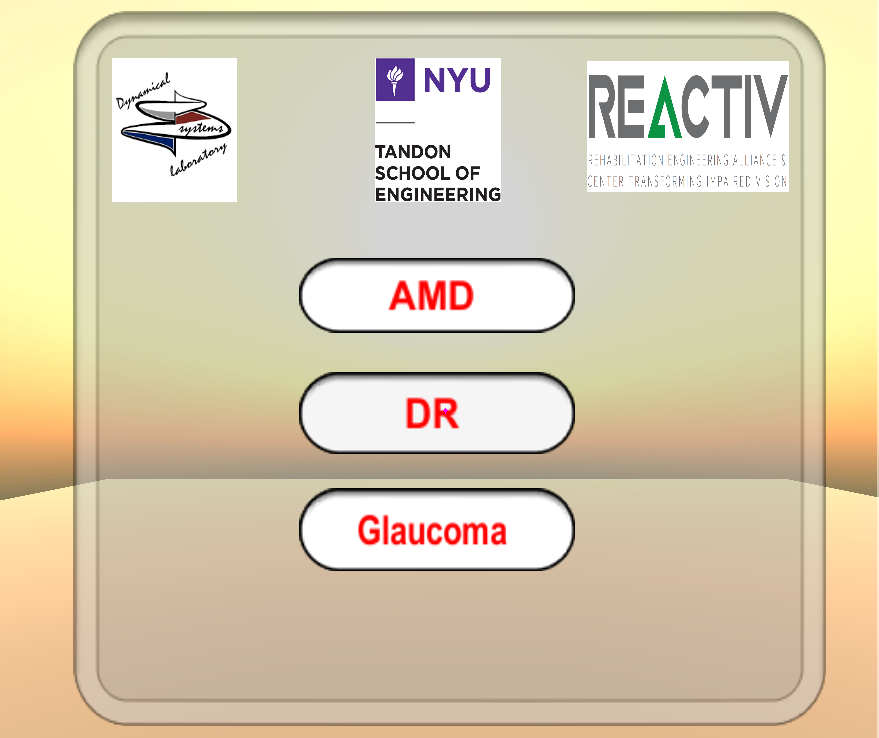}  & \includegraphics[width=0.49\textwidth]{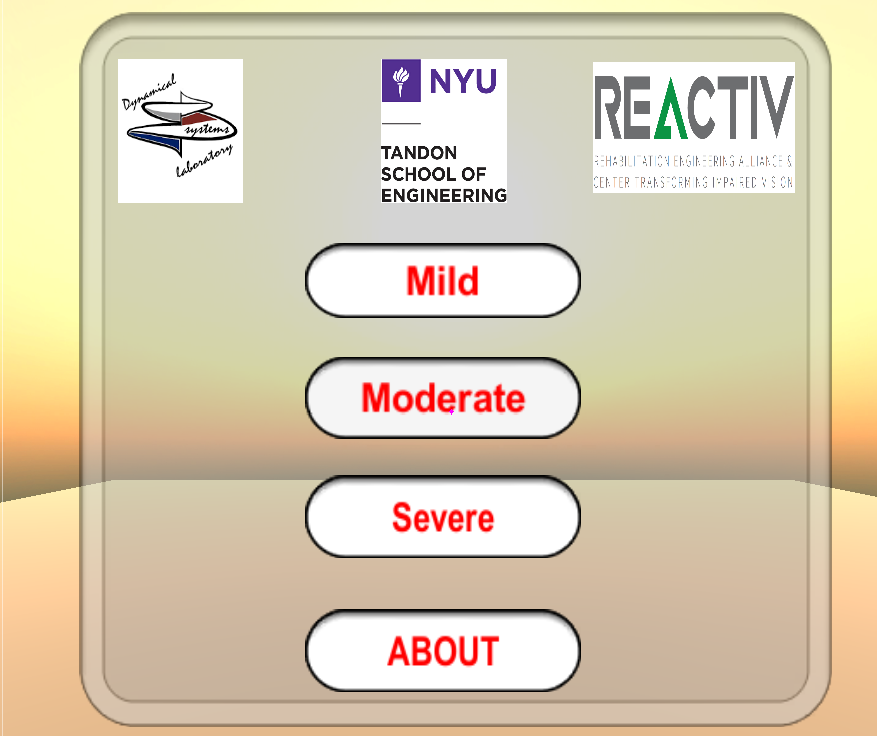} \\
\end{tabular}
\begin{tabular}{c}
    (c) \\  \includegraphics[width=0.49\textwidth]{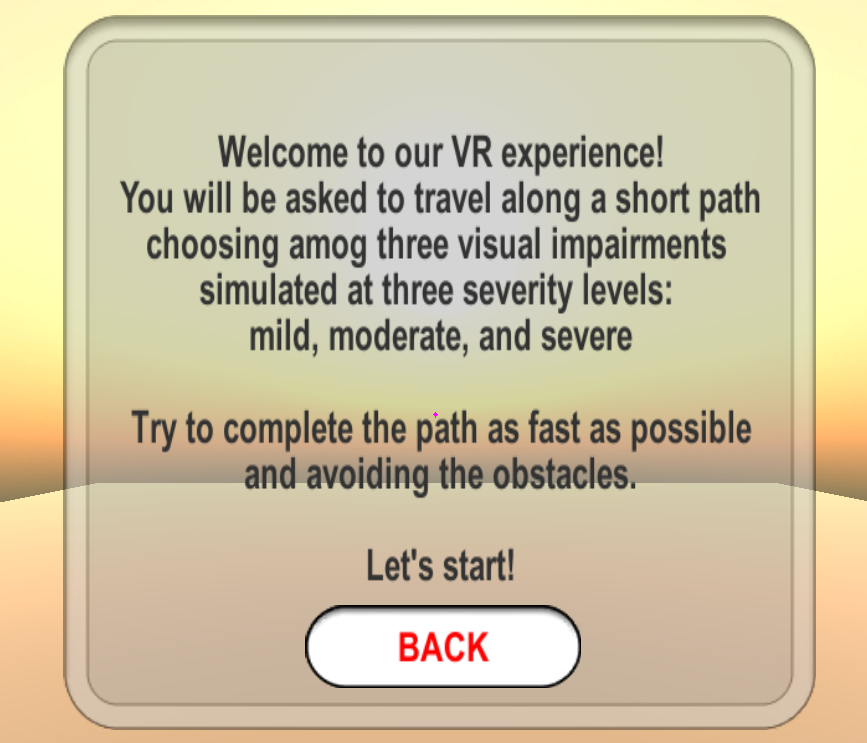} \\
\end{tabular}
\caption{Example of menu screens used for: (a) selecting the vision loss type (AMD, DR, Glaucoma), (b) selecting the severity stage (Mild, Moderate, Severe) or the \textit{ABOUT} section, and (c) accessing the \textit{ABOUT} section providing instructions and information about the VR experience.}   
\label{fig:fig6}
\end{figure}

To facilitate the experience, the room was set up so that each participant had their own VR headset connected to a PC (see Fig. \ref{fig:fig5}b). This arrangement allowed the experience to be streamed on a screen, enabling us to assist users while they performed tasks, if necessary. 
In the VR environment, participants could choose from a menu which VL they would like to experience--AMD, DR, or Glaucoma--and select the severity of the impairment: mild, moderate, or severe (see Fig. \ref{fig:fig6}a-b). Additionally, an \textit{ABOUT} button in the menu projected a canvas with information about the experience and tasks (see Fig. \ref{fig:fig5}c).

Participants were tasked with navigating through the virtual urban environment, attempting to avoid obstacles and complete the task as quickly as possible. They were encouraged to try as many VLs and stages as they wished and were informed that they could stop the experience at any time if they felt discomfort, tiredness, or for any other reason.

\subsection{Questionnaire and focus group}
The NYU-DOT InSight workshop incorporated comprehensive pre- and post-intervention questionnaires to gauge the effectiveness of the VR experience in raising awareness about the challenges faced by persons with VL. The pre-intervention questionnaire aimed to assess participants' baseline knowledge and attitudes towards blindness, low vision, and related design guidelines (see Tab. \ref{tab:tab1}). Questions focused on their confidence in defining blindness and low vision, awareness of mobility issues, and familiarity with virtual reality technology as a tool for testing design guidelines. 
\begin{table}[!h]
\caption{InSight workshop – pre-intervention questionnaire}
\begin{tabular}{|p{0.20\textwidth}|c|c|c|c|c|}
\hline
\textbf{Question} & \textbf{1 (Not at all)} & \textbf{2 (Slightly)} & \textbf{3 (Moderate)} & \textbf{4 (Quite)} & \textbf{5 (Extremely)} \\ \hline
How confident are you with the definitions of blindness and low vision? & \multicolumn{1}{c|}{$\Box$} & \multicolumn{1}{c|}{$\Box$} & \multicolumn{1}{c|}{$\Box$} & \multicolumn{1}{c|}{$\Box$} & \multicolumn{1}{c|}{$\Box$} \\ \hline
How much are you aware about design guidelines of spaces for persons with blindness or low vision? & \multicolumn{1}{c|}{$\Box$} & \multicolumn{1}{c|}{$\Box$} & \multicolumn{1}{c|}{$\Box$} & \multicolumn{1}{c|}{$\Box$} & \multicolumn{1}{c|}{$\Box$} \\ \hline
How confident are you about mobility issues of persons with blindness or low vision? & \multicolumn{1}{c|}{$\Box$} & \multicolumn{1}{c|}{$\Box$} & \multicolumn{1}{c|}{$\Box$} & \multicolumn{1}{c|}{$\Box$} & \multicolumn{1}{c|}{$\Box$} \\ \hline
Do you consider persons with visual impairment needs when designing spaces? & \multicolumn{1}{c|}{$\Box$} & \multicolumn{1}{c|}{$\Box$} & \multicolumn{1}{c|}{$\Box$} & \multicolumn{1}{c|}{$\Box$} & \multicolumn{1}{c|}{$\Box$} \\ \hline
How much are you aware about virtual reality technology? & \multicolumn{1}{c|}{$\Box$} & \multicolumn{1}{c|}{$\Box$} & \multicolumn{1}{c|}{$\Box$} & \multicolumn{1}{c|}{$\Box$} & \multicolumn{1}{c|}{$\Box$} \\ \hline
How much are you aware about the use of virtual reality technology for testing design guidelines of spaces for persons with blindness or low vision? & \multicolumn{1}{c|}{$\Box$} & \multicolumn{1}{c|}{$\Box$} & \multicolumn{1}{c|}{$\Box$} & \multicolumn{1}{c|}{$\Box$} & \multicolumn{1}{c|}{$\Box$} \\ \hline
Do you think that a virtual reality experience could represent an innovative and immersive way for increasing awareness of blindness and low vision? & \multicolumn{1}{c|}{$\Box$} & \multicolumn{1}{c|}{$\Box$} & \multicolumn{1}{c|}{$\Box$} & \multicolumn{1}{c|}{$\Box$} & \multicolumn{1}{c|}{$\Box$} \\ \hline
\end{tabular}
\label{tab:tab1}
\end{table}

The post-intervention questionnaire, administered after the VR experience, evaluated changes in participant knowledge, confidence, and perceptions  (see Tab. \ref{tab:tab2}). It included questions on the usefulness of the program, willingness to discuss the experience with others, and interest in future similar programs. By comparing responses from both questionnaires, we can discern the impact of the VR experience on participant understanding and attitudes towards the needs of persons with VL. These insights are crucial for informing future initiatives and enhancing the inclusivity of urban environments. 

\begin{table}[!h]
\caption{InSight workshop – post-intervention Questionnaire}
\begin{tabular}{|p{0.20\textwidth}|c|c|c|c|c|}
\hline
\textbf{Question} & \textbf{1 (Not at all)} & \textbf{2 (Slightly)} & \textbf{3 (Moderate)} & \textbf{4 (Quite)} & \textbf{5 (Extremely)} \\ \hline
How confident are you with the definitions of blindness and low vision? & \multicolumn{1}{c|}{$\Box$} & \multicolumn{1}{c|}{$\Box$} & \multicolumn{1}{c|}{$\Box$} & \multicolumn{1}{c|}{$\Box$} & \multicolumn{1}{c|}{$\Box$} \\ \hline
How much are you aware about design guidelines of spaces for persons with blindness or low vision? & \multicolumn{1}{c|}{$\Box$} & \multicolumn{1}{c|}{$\Box$} & \multicolumn{1}{c|}{$\Box$} & \multicolumn{1}{c|}{$\Box$} & \multicolumn{1}{c|}{$\Box$} \\ \hline
How confident are you about mobility issues of persons with blindness or low vision? & \multicolumn{1}{c|}{$\Box$} & \multicolumn{1}{c|}{$\Box$} & \multicolumn{1}{c|}{$\Box$} & \multicolumn{1}{c|}{$\Box$} & \multicolumn{1}{c|}{$\Box$} \\ \hline
Would you consider what you learnt in your future work? & \multicolumn{1}{c|}{$\Box$} & \multicolumn{1}{c|}{$\Box$} & \multicolumn{1}{c|}{$\Box$} & \multicolumn{1}{c|}{$\Box$} & \multicolumn{1}{c|}{$\Box$} \\ \hline
How much are you aware about virtual reality technology? & \multicolumn{1}{c|}{$\Box$} & \multicolumn{1}{c|}{$\Box$} & \multicolumn{1}{c|}{$\Box$} & \multicolumn{1}{c|}{$\Box$} & \multicolumn{1}{c|}{$\Box$} \\ \hline
How much are you aware about the use of virtual reality technology for training mobility of persons with blindness or low vision? & \multicolumn{1}{c|}{$\Box$} & \multicolumn{1}{c|}{$\Box$} & \multicolumn{1}{c|}{$\Box$} & \multicolumn{1}{c|}{$\Box$} & \multicolumn{1}{c|}{$\Box$} \\ \hline
Do you think that a virtual reality experience could represent an innovative and immersive way for increasing awareness of blindness and low vision? & \multicolumn{1}{c|}{$\Box$} & \multicolumn{1}{c|}{$\Box$} & \multicolumn{1}{c|}{$\Box$} & \multicolumn{1}{c|}{$\Box$} & \multicolumn{1}{c|}{$\Box$} \\ \hline
How useful did you find this program? & \multicolumn{1}{c|}{$\Box$} & \multicolumn{1}{c|}{$\Box$} & \multicolumn{1}{c|}{$\Box$} & \multicolumn{1}{c|}{$\Box$} & \multicolumn{1}{c|}{$\Box$} \\ \hline
Do you think you will talk about this program with other people? & \multicolumn{1}{c|}{$\Box$} & \multicolumn{1}{c|}{$\Box$} & \multicolumn{1}{c|}{$\Box$} & \multicolumn{1}{c|}{$\Box$} & \multicolumn{1}{c|}{$\Box$} \\ \hline
Do you think you will be interested in being involved in another program like this? & \multicolumn{1}{c|}{$\Box$} & \multicolumn{1}{c|}{$\Box$} & \multicolumn{1}{c|}{$\Box$} & \multicolumn{1}{c|}{$\Box$} & \multicolumn{1}{c|}{$\Box$} \\ \hline
\end{tabular}
\label{tab:tab2}
\end{table}

A few days after the workshop, a focus group was held to gather more detailed feedback from participants. The timing allowed them to reflect on the VR experience, offering a more thoughtful evaluation of both the immersive simulation and its impact on their understanding of the challenges faced by persons with VL in urban environments.

The focus group explored how participant perspectives had evolved after having time to process the experience. They discussed the realism of the VR scenarios, the effectiveness of the simulation in raising awareness about mobility challenges, and the potential of VR as a tool for training and education. Participants reflected on the emotional and cognitive impact of navigating virtual environments with simulated VLs and shared insights on how this influenced their empathy toward persons with these conditions.

Feedback also covered suggestions for improving the VR experience, from addressing technical issues to enhancing the realism of the scenarios. Participants discussed the broader application of the technology, considering how it could be used in future training programs or to raise awareness among different audiences, such as public officials and urban planners, to promote inclusive design in public spaces.

Ultimately, the focus group provided valuable insights into how the VR workshop influenced participant thinking, with many expressing a greater awareness of the needs of persons with VL. This feedback will guide future improvements to ensure the VR experience continues to promote inclusivity and accessibility in urban design.

\subsection{Interviews}
Members of the Vision Zero Task Force, a multidisciplinary team established by the Vision Zero Action Plan, were interviewed to gather insights into ongoing and future traffic safety initiatives. This team consists of experts from across various NYC agencies, all dedicated to improving traffic safety. Their role involves reviewing and coordinating ongoing and future projects, assessing progress on safety measures, and fostering collaboration across different governmental bodies. 

We specifically interviewed this team due to their central role in planning and executing initiatives aimed at enhancing pedestrian safety, which directly aligns with our project goal of improving the mobility and safety of vulnerable populations. Their expertise in traffic safety, combined with their experience in implementing citywide projects, made them ideal contributors to our understanding of how existing and future interventions could be optimized for accessibility.

The interviews were structured around questions related to key DOT programs, such as the extension of Leading Pedestrian Intervals (LPIs), the Turn Calming Program, and Street Improvement Projects (SIPs), all of which are essential in creating safer streets, especially for seniors and persons with VL. Additionally, we explored how DOT integrates accessible pedestrian signals (APS) and curb ramp improvements to enhance independent mobility \cite{WBG1}. The focus was also on how current plans, like the Bus Stops Under the Elevated (BSUE) initiative, aim to make bus stops more accessible and predictable for pedestrians, including those with mobility and VL.

The questions asked in the interview (see Interview S1 in the Supplemental Information) were based on the detailed action plans and strategies outlined in \textquoteleft{Pedestrian Safety and Older New Yorkers}\textquoteright{} report \cite{PedestrianSafety}. This report emphasizes reducing hazard exposure and improving safety through infrastructure changes and community outreach efforts. The interviews sought to clarify the progress, timeline, and specific solutions that DOT is committed to implementing across the city, ensuring safer navigation and mobility for all pedestrians, particularly the most vulnerable.

\section{Results}
\subsection{Pre- and post-questionnaire analysis}

The pre- and post-questionnaire data indicate that the VR intervention significantly improved participants understanding and awareness of the challenges faced by persons with VL in urban environments. Prior to the VR simulation, only 10\% of participants felt highly confident in their understanding of key terms such as blindness and low vision. However, post-questionnaire results revealed that 65\% of participants rated themselves as \textquoteleft{quite}\textquoteright{} or \textquoteleft{extremely}\textquoteright{} confident, illustrating that the immersive nature of VR effectively deepened comprehension. This finding aligns with previous research, which highlights VR's ability to make abstract or unfamiliar concepts more tangible by allowing users to experience these challenges firsthand in a controlled, immersive environment \cite{muller2022traveling,villalba2021state}.

A similar trend was observed in participants' awareness of urban design guidelines affecting persons with VL. Initially, fewer than 15\% of participants rated themselves as highly aware of how urban design impacts the mobility of persons with VL. After the intervention, over 50\% of respondents indicated a significantly higher level of awareness. By interacting with the virtual environment, participants better understood how design elements such as tactile paving and auditory cues either aid or hinder navigation for persons with VL. This echoes the findings of Giudice et al. \cite{giudice2008blind}, who demonstrated that VR offers unique advantages in educating users about urban accessibility by immersing them in virtual environments that simulate real-world design challenges.

Moreover, participants demonstrated a substantial increase in confidence when discussing mobility issues related to VL. Prior to the intervention, fewer than 20\% of participants felt highly confident in their understanding of mobility challenges, whereas post-intervention, nearly 70\% felt confident. The VR simulation allowed participants to experience real-world challenges such as uneven terrain and environmental obstacles, helping to bridge the gap between theoretical knowledge and practical application. This approach, which is central to VR's educational value, is supported by previous work that recognizes VR as a tool for experiential learning \cite{riva2005virtual}.

Participant familiarity with the use of VR for urban design testing also increased significantly. Before the intervention, only 25\% of participants were aware of VR's potential for testing urban design guidelines. After the intervention, nearly 80\% recognized VR as a valuable tool for evaluating the accessibility of urban spaces. This reflects growing recognition of VR as a means to trial urban design concepts in a risk-free, cost-effective manner, thereby enhancing planning processes and improving outcomes \cite{winters2010route}.

Additionally, over 80\% of participants expressed that they would now consider the needs of persons with VL in their future designs, compared to 40\% before the intervention. This increase suggests that the VR experience not only raised awareness but also fostered empathy and motivated participants to implement more inclusive practices. Prior research has shown that VR-based immersive experiences can inspire long-term behavioral changes by allowing users to "live" through the challenges of others, which was similarly observed in this study \cite{slater2009place}.

\subsection{Focus group discussion}

The focus group further corroborated the findings from the questionnaire, emphasizing the effectiveness of VR in raising awareness, acceptance, and understanding of the challenges faced by persons with VL. Participants praised the immersive nature of the workshop, which allowed them to see firsthand how persons with VL navigate urban environments. One participant highlighted that the VR simulation provided a safe, controlled, and highly engaging environment for conveying the impact of vision loss, significantly contributing to building empathy and awareness \cite{yao2021evaluating,pinto2023inclusion}.

Several participants also noted the practical implications of the VR workshop, particularly in testing new urban design concepts. They appreciated the ability to trial street markings, traffic calming measures, and accessibility features in a virtual environment, which they recognized as both cost-effective and efficient compared to physical modifications to the built environment \cite{van2018effectiveness,gordon2011augmented,imottesjo2022urban}. Many participants expressed enthusiasm about the potential for integrating VR simulations with real-world urban planning processes, especially for evaluating designs involving complex interactions between freight vehicles, micro-mobility, and pedestrians \cite{van2018effectiveness,gordon2011augmented}.

One of the key takeaways was the ability of VR to exceed participant expectations, broadening their understanding of urban mobility challenges beyond the needs of persons with VL. Participants felt that the VR experience offered more comprehensive insights into urban design and expressed a desire to incorporate these insights into their future work. Some suggestions for improvement included enhancing the realism of the simulation by incorporating more diverse elements, such as trucks with blind spots, obstacles like food carts, and micro-mobility vehicles. These additions would make the experience even more relevant and practical for urban planners and designers \cite{oing2018implementations,maples2017use}.

The application of the workshop's lessons to real-world urban planning was particularly appreciated. Participants recognized that the VR simulation allowed them to test design decisions and assess their impacts on persons with VL before implementing these changes in the real world. This approach provides a valuable tool for creating more accessible and inclusive urban spaces. As noted in previous studies \cite{krull2017research}, the capacity of VR to simulate and evaluate urban environments can significantly enhance the planning process by allowing for adjustments to be made in a controlled, virtual setting.

\section{Discussion}
This study explored the potential of VR technology to improve design considerations of urban environments in an effort to enhance accessibility. The study involved DOT participants, who engaged in a VR intervention simulating various visual impairments within urban settings. This immersive experience allowed them to gain firsthand insights into the mobility challenges that persons with VL encounter, such as navigating poorly designed urban spaces. The methodology included the creation of a custom VR simulating a diversity of VLs, including various severities. Participants were asked to navigate virtual environments designed to emulate real-world obstacles that persons with VL commonly face, such as insufficient tactile feedback, poor contrast in wayfinding materials, and improperly marked crossings. This provided a unique, hands-on experience, distinguishing this study from other approaches relying solely on theoretical or observational learning. 

Empirical findings indicate significant gains in participant awareness, empathy, and confidence in applying inclusive design principles. However, beyond reporting these findings, it is important to analyze the broader implications of this study and connect them to relevant themes emerging from existing literature. The results of this study highlight three key themes: i) programmatic growth through deeper understanding, ii) innovative methodologies for urban planning, and iii) potential future directions for the role of VR in urban planning and disability equity. These themes emerged as participants not only gained valuable insights into the daily challenges faced by individuals with visual disabilities but also began to think critically about how to incorporate these insights into more inclusive design strategies. 

\subsection{Programmatic growth through deeper understanding}
A major theme emerging from the study is the potential for broader programmatic growth through deeper understanding of the mobility challenges faced by persons VL. Insights from interviews with the Vision Zero Task Force—a team dedicated to traffic safety initiatives—highlighted how VR technology could enhance ongoing DOT programs aimed at improving pedestrian safety for vulnerable populations.

Key initiatives such as Extended Leading Pedestrian Intervals (LPIs), which increase crossing times, and the Turn Calming Program, which slows vehicle speeds at intersections, are vital for enhancing accessibility. VR simulations could provide a cost-effective way to test these interventions, ensuring they adequately address the needs of persons with VL. This aligns with DOT members' views on refining tactile paving and auditory cues by trialing them in virtual environments before real-world implementation.

Additionally, DOT’s broader Street Improvement Projects (SIPs), including road diets and pedestrian islands, emphasize infrastructure changes that improve safety. Programs like Accessible Pedestrian Signals (APS) and curb ramp upgrades further support navigation for persons with VL. VR can be used to simulate SIPs in realistic settings, allowing for adjustments that maximize effectiveness.

The Bus Stops Under the Elevated (BSUE) Program, designed to improve bus stop accessibility, was another key focus. DOT officials noted that VR could be used to test and refine designs, particularly in challenging environments like those under elevated train structures, where poor visibility and complex layouts are common.

Incorporating VR into these initiatives provides an opportunity to preemptively assess and enhance design solutions, ensuring they meet the needs of pedestrians with VL. By integrating immersive VR simulations into existing programs, the DOT can improve safety outcomes while fostering long-term growth in inclusive urban planning. Additionally, we could integrate a trucks-eye-view simulation, replicating the view from a truck driver's perspective, into VR programs. This could offer unique insights into how freight vehicles interact with urban infrastructure. By simulating complex maneuvers like turning at tight intersections or navigating through areas with heavy pedestrian traffic, the DOT could identify design challenges and optimize streets for safer and more efficient freight movement, while also improving pedestrian and cyclist safety.

\subsection{Innovative methodologies for urban planning}
In addition to enhancing understanding, this study demonstrates the promise of VR as an innovative methodology for evaluating and testing urban design solutions. Participants widely recognized VR as a cost-effective and efficient tool for simulating real-world environments and testing design changes without physical modifications. This finding is particularly significant when considering future directions in urban planning, where design solutions for accessibility can be trialed and refined in virtual environments before implementation.

Current VR applications used in this study simulate vision impairments ranging from mild to severe. However, future iterations could expand upon this by introducing a continuum of visual impairments, allowing for more nuanced testing of design elements under varied conditions. As urban environments become increasingly complex with the integration of micro-mobility vehicles and evolving street designs, VR simulations must keep pace by incorporating these dynamic interactions. This was echoed by participants during focus group discussions, during which suggestions were made to incorporate additional elements such as trucks with blind spots, food carts, and scooters to increase the realism and simulation applicability.

Moreover, this study highlights the potential for VR to exacerbate underlying conditions in simulated environments to better prepare both urban planners and persons with VL for future challenges. For example, VR could simulate worsening visibility conditions due to weather or the gradual loss of vision over time, helping to assess whether urban spaces remain accessible as conditions change. Such advanced simulations could provide valuable data for predicting how current infrastructure might fail persons with VL in different scenarios, thereby informing more resilient and forward-thinking designs.

\subsection{Future directions: moving beyond the present approach}
While the VR intervention focused on simulating specific levels of VL, there is a growing need to move beyond static categories (mild, moderate, and severe) and toward a more fluid continuum of disabilities. This shift aligns with broader discussions in accessibility research, which advocate for more individualized approaches to designing for persons with disabilities. Future VR applications could allow participants to experience the gradual deterioration of vision over time, highlighting the need for adaptive urban design features that can support pedestrians at any stage of their vision-loss journey.

VR environments could be enhanced by introducing dynamic factors, such as changing ambient conditions. For example, simulations could alter the time of day to show how varying light levels—such as daylight, dusk, or night—affect visibility and navigation for persons with VL. Similarly, incorporating weather conditions like rain, fog, or snow could demonstrate how precipitation and reduced visibility impact accessibility. These variable settings would allow urban planners and designers to explore how different environmental factors interact with VLs, and adjust designs to account for these real-world challenges.

As cities continue to adopt new technologies in urban mobility—such as autonomous vehicles, advanced pedestrian detection systems, and integrated data-sharing between road users—VR can serve as a testing ground for how these innovations impact persons with VL. By integrating VR simulations with real-world urban planning initiatives like the DOT's Extended LPIs and Turn-Calming Program, urban designers could better anticipate how new technologies and street designs might interact with various disabilities, ensuring that future urban environments are inclusive and adaptive to changing needs. There is also the potential to incorporate an artificial intelligence (AI) dimension into VR simulations. AI could simulate dynamic pedestrian and vehicle behaviors in real time, creating a more realistic and responsive urban environment for testing. These AI-driven simulations could adapt to evolving mobility patterns, such as predicting how a person with deteriorating vision might navigate urban spaces or how autonomous vehicles would interact with pedestrians. This would allow for more accurate assessments of urban design features and their impacts on persons with varying levels of VL.

\section{Conclusion}
In conclusion, this study demonstrates that VR technology holds significant promise for enhancing understanding, empathy, and innovation in urban planning for persons with VL. By immersing participants in the lived experiences of pedestrians with VL, VR not only deepened their understanding but also provided a practical tool for evaluating and refining urban design solutions. Programmatic growth in urban planning departments can be driven by the integration of VR simulations into existing initiatives, while future directions should focus on expanding capabilities of VR to simulate a continuum of VL and complex urban interactions.

The findings underscore the potential of VR to serve as a catalyst for systemic change in urban planning, equipping designers and policymakers with the tools necessary to create more accessible, resilient, and inclusive urban spaces. By continuing to push the boundaries of what VR can simulate, future research and practice will help ensure that urban environments evolve to meet the needs of all pedestrians, particularly those with visual disabilities.

\backmatter

\bmhead{Supplementary information}
The questions used during the interviews with DOT members are detailed in the supplementary file Interview S1

\bmhead{Acknowledgements}
The authors would like to express their gratitude to Denise Ramirez and the leadership of Transportation Planning and Management Division of the New York City Department of Transportation. Their assistance was invaluable in enabling us to organize and conduct the VR-based workshop. We appreciate their commitment to advancing accessibility and inclusive design in urban planning, which was essential to the success of this study. 

\section*{Declarations}
\bmhead{Funding}
This study was supported by the National Science Foundation under award number ECCS-1928614, DUE-2129076, CNS-1952180, ITE-2236097, and ITE-2345139.

\bmhead{Conflicts of interests}
The authors have no conflicts of interest to declare that are relevant to the content of this article

\bibliography{sn-bibliography}% common bib file

\end{document}